# Reduced nonlinear description of Farley-Buneman instability

A. S. Volokitin[1] , B. Atamaniuk[2]

[1] Institute of Terrestrial Magnetism, the Ionosphere, and Radio Wave Propagation, RAN, IZMIRAN, 142092 Troitsk, Moscow Region, Russia .

[2] Institute of Fundamental Technological Research PAS (IFTR PAS) Świętokrzyska 21, 00-049 Warszawa, POLAND

**Abstract**. In the study on nonlinear wave-wave processes in an ionosphere and a magnetosphere usually the main attention is paid to investigation of plasma turbulence at well developed stage, when the wide spectrum of plasma wave is present. On the other side, it is well known that even if the number of cooperating waves remains small due to a competition of processes of their instability and attenuation, the turbulence appears in the result of their stochastic behavior. The regimes of nonlinear dynamics of low frequency waves excited due to Farley-Buneman instability in weakly ionized and inhomogeneous ionospheric plasma in the presence of electric current perpendicular to ambient magnetic field are considered. The problem is essentially three dimensional and difficult for full numerical simulation, but the strong collisional damping of waves allow to assume that in this case a perturbed state of plasma can be described as finite set of interacting waves, some of which are unstable and other strongly damping. The proposed nonlinear model allow to make full study of nonlinear stabilization, conditions of stochasticity and to consider the different regimes and properties of few mode plasma turbulence.

## 1. Introduction

The study on nonlinear wave-wave interaction is an essential part of researches of wave processes in an ionosphere and a magnetosphere. Usually the main attention is paid to investigation of plasma turbulence at well developed stage, when the wide spectrum of plasma wave is present. On the other side, it is well known that even system with finite number of interacting waves can realize a turbulent state in active media. In such cases, when the number of cooperating waves remains small due to a competition of processes of their instability and attenuation, the turbulence appears in the result of their stochastic behavior. The perturbed ionospheric plasma is one of important example of such active media. Owing to the complicated interaction of a solar wind with the Earth magnetosphere an electrojet and corresponding DC electric field arise in E- region of the lower ionosphere. If value of this field exceeds some threshold the Farley-Buneman (FB) instability collisional plasmas with an electric current across a magnetic field develops. This instability was discovered more than 30 years ago [1]and is its linear theory [1], [2], [3] allow to explain a number of observable properties of small-scale inhomogeneities arising in the inferior ionosphere. Thus it predict that the short waves spread almost perpendicularly to an exterior magnetic field are most promptly raised. However to explain a depth of modulation and other properties of a spectrum of density inhomogeneities it is necessary to know the mechanism of the instability stabilization. In number of works is proposed that FB instability is saturated due to non-linear interaction of instable modes with decaying mode and the concrete mechanism of nonlinearity has been offered. As the characteristic times of oscillations of density exceed a cyclotron period the vector nonlinearity caused by nonlinearity of a drift motion of electrons plays defining role in an establishment of a spectrum of inhomogeneities of plasma or electrostatic waves in weakly collisional plasma [4], [5], [6]. Consequently, the main nonlinearity results from perpendicular to magnetic field convection of a density perturbation in one wave by another wave and mathematically is expressed in a specific vector form. The problem is essentially three dimensional. Moreover it is necessary to take into account such essential effects as Landau damping on ions, but the full kinetic modeling encounters obvious difficulties: shortage of computer resources, complexities of the adequate description of collisions, presence of numerical noise, a major difference in temporary gauges of oscillations of particles and velocity of non-linear processes.

At essential a role of the dissipative processes allow to suggest, that the developing spectrum (turbulence) of oscillations of density can be presented limited number of modes. The dynamic of this set of waves now can be described by mathematical system of ordinary differential equations, which describe the developing few-mode turbulence. This system can be effectively featured by means of combined equations of magnetohydrodynamic type for amplitudes of oscillation which take into account effects of Landau damping due to resonant interactions with ions. This allow to avoid difficulties of 3D simulations and to make full study of nonlinear stabilization, conditions of stochasticity and to consider the different regimes and properties of few plasma turbulence in the conditions when the number of interacting waves keeps small by the strong competition of processes wave damping and instabilities.

## 2. The Farley-Buneman instability

The linear theory of instability Farley-Buneman was considered nonsinglely and with use of various models of the description of the magnetized plasma, the using both magneto-hydrodynamic and kinetic description. In this section we, not pretending for novelty of the analysis, will give the basic effects of the linear theory which will be used at build-up of non-linear model. In viewed requirements it is possible to consider plasma as homogeneous then in the linear approach of property of electrostatic oscillations of plasma are defined from the dispersion equation $1+\varepsilon_e(\omega,\mathbf{k})+\varepsilon_i(\omega,\mathbf{k})=0$, where $\varepsilon_e(\omega,\mathbf{k})$ and $\varepsilon_i(\omega,\mathbf{k})$ are dielectric permittivity of electrons and ions correspondingly. It is reasonable to consider evolution of density perturbations in the reference frame where ions are rest. Owing to the strong friction due to high frequency $\nu_i$ collisions of ion with neutral plasma component this reference frame practically coincides with system of rest of neutral component of weakly ionized ionosphere plasmas. Considering plasma density oscillation with frequency of $\omega$ much less then electron cyclotron frequency $\omega_e = eB/m_e c$, $\omega \ll \omega_e$, we suppose that electrons are magnetized and as whole drift with velocity $\mathbf{v}_d = \mathbf{v}_D \equiv \frac{c}{\mathbf{B}^2}[\mathbf{E}_0,\mathbf{B}]$ relatively ions under the action of constant electric field $\mathbf{E}_0$. Then using electron part of an permittivity in collisional plasmas (see [7]), supposing also that $k_\| v_{Te} \ll \omega_e, \nu_e$, $|\omega - \mathbf{k}\mathbf{v}_d| \ll \nu_e$ and $k_\perp^2 v_{Te}^2 \ll \omega_e^2$ and taking into account that $\nu_e \gg |\omega - \mathbf{k}\mathbf{v}_d|$ we get the next approximate expression for $\varepsilon_e$

$$\varepsilon_e \simeq \frac{\omega_p^2}{\omega_e^2} \frac{(\omega - \mathbf{k}\mathbf{v}_d + i\nu_e)\left(\frac{k_\perp^2}{k^2} + \frac{k_\|^2}{k^2}\frac{\omega_e^2}{\nu_e^2}\right)}{\omega - \mathbf{k}\mathbf{v}_d + i\nu_e\left(\frac{k_\perp^2 v_{Te}^2}{\omega_e^2} + \frac{k_\|^2 v_{Te}^2}{\nu_e^2}\right)} \simeq \frac{\omega_p^2}{\omega_e^2} \frac{i\nu_e\left(\frac{k_\perp^2}{k^2} + \frac{k_\|^2}{k^2}\frac{\omega_e^2}{\nu_e^2}\right)}{\omega - \mathbf{k}\mathbf{v}_d + i\nu_e\left(\frac{k_\perp^2 v_{Te}^2}{\omega_e^2} + \frac{k_\|^2 v_{Te}^2}{\nu_e^2}\right)} \qquad (1)$$

where $n_0$ - plasma density, $\omega_p^2 = \frac{4\pi e^2 n_0}{m_e}$, $\lambda_D^2 = T_e 4\pi e^2 n_0$, $v_{T_{i,e}} = \sqrt{T_{i,e} m_{i,e}}$ - thermal velocities and $T_i, T_e$ - temperatures of ions and electrons. In the same way using the general expressions for an permittivity in collisional plasmas [7] in case of not magnetized ions ($\omega \gg \omega_i = \frac{eB_0}{m_i c}$), supposing also a small thermal velocity of ions $v_{T_i}$ in comparison with phase velocity of oscillations, $|\omega + i\nu_i| \gg kv_{T_i}$, get

$$\varepsilon_i \simeq -\frac{\omega_{pi}^2}{(\omega + i\nu_i)\omega}\left[1 - i\sqrt{\frac{\pi}{2}}\left(\frac{\omega + i\nu_i}{kv_{T_i}}\right)^3 \exp\left[-(\omega + i\nu_i \sqrt{2} kv_{T_i})^2\right]\right] \quad (2)$$

In the case of long wave oscillations it is possible to neglect ion Landau damping and from $\varepsilon_e(\omega, \mathbf{k}) + \varepsilon_i(\omega, \mathbf{k}) \simeq 0$ follows the dispersion equation of FB oscillations (see below) solution of which describe quasineutral plasma oscillations with frequency and growth rate

$$Re\omega = \Omega_k \simeq \frac{\mathbf{k v}_d}{1 + \Psi_0}, Im\omega = \gamma_{FB} \simeq \nu_e\left(\frac{k_\perp^2}{k^2} + \frac{k_\parallel^2}{k^2}\frac{\omega_e^2}{\nu_e^2}\right)\frac{\omega^2 - c_s^2 k^2}{\omega_{lh}^2 (1 + \Psi_0)} \quad (3)$$

where $\omega_{lh}^2 = \omega_{pi}^2 \omega_e^2 / \omega_p^2$, $c_e^2 = \frac{T_e}{m_i}$, $c_s^2 = \frac{5}{3}\frac{T_e + T_i}{m_i}$ and $\Psi_0 = \frac{\nu_e \nu_i}{\omega_{lh}^2}\left(\frac{\omega_e^2}{\nu_e^2}\frac{k_z^2}{k_\perp^2} + 1\right)$. Note that oscillations are unstable if $\omega^2 \geq c_s^2 k^2$ and $\mathbf{k v}_d > kc_s(1 + \Psi_0)$ and growth rate $\gamma_{FB}$ is increasing with increasing of wave number $k$ as $\gamma_{FB} \sim k^2$. However as phase velocity of oscillations $v_{ph} = \Omega_k / k = \frac{v_d}{1+\Psi_0}\frac{k_x}{k} = \frac{v_d}{1+\Psi_0}\cos\varphi$ appears the order of thermal velocity of ions, resonant interaction of ions became essential and bring stabilization of instability at short wavelengths [?], [?] due to Landau damping on ions. The effect of Landau damping require the kinetic description, but within the limits of the linear approach it is enough to retain an imaginary part $\varepsilon_i$. Then, defining $\chi = \sqrt{\frac{\pi}{2}}\left(\frac{\omega + i\nu_i}{kv_{T_i}}\right)^3 \exp\left[-(\omega + i\nu_i \sqrt{2} kv_{T_i})^2\right]$ and omitting at $\omega_e^2 \ll \omega_p^2$ small terms which responsible for the violation of the quasi-neutrality, we can present the dispersion equation of quasineutral FB oscillations in the presence of Landau damping

$$\omega = \frac{\mathbf{k v}_d(1 - i\chi) + i\nu_e\left(\frac{k_\perp^2}{k^2} + \frac{k_\parallel^2}{k^2}\frac{\omega_e^2}{\nu_e^2}\right)\frac{\omega^2 - (1-\chi)c_s^2 k^2}{\omega_{lh}^2}}{1 - i\chi + \frac{\nu_e \nu_i}{\omega_{lh}^2}\left(\frac{k_\perp^2}{k^2} + \frac{k_\parallel^2}{k^2}\frac{\omega_e^2}{\nu_e^2}\right)}. \quad (4)$$

If $|\chi| \ll 1$, the influence of kinetic effects on frequency of FB oscillation is small, $Re\omega \gg Im\omega$, and is possible to see that frequency of oscillation varies negligible in comparison with (3) $Re\omega \simeq \frac{\mathbf{k v}_d(1+Im\chi)}{1+\Psi_0+Im\chi}$, and resulting growth rate can be presented as sum of classic expression and ion Landau damping

$$\gamma = Im\omega \simeq \frac{\mathbf{k v}_d(-Re\chi) + \frac{Re\chi}{1+\Psi_0+Im\chi} + \nu_e\left(\frac{k_\perp^2}{k^2} + \frac{k_\parallel^2}{k^2}\frac{\omega_e^2}{\nu_e^2}\right)\frac{\omega^2 - \omega_{lh}^2 k^2 v_{Te}^2}{\omega_{lh}^2}}{1 + \Psi_0 + Im\chi} \simeq \gamma_{FB} - Re\omega Re\chi \quad (5)$$

## 3. Nonlinear model in approximation of the weak interaction

To build a model suitable for study a few mode turbulence in the ionosphere we consider low-frequency electrostatic oscillations in collisional plasma at $v_i \leq \frac{\partial}{\partial t} \ll v_e$, which can be described by the standard equations of two-fluid magnetohydrodynamics for density of electrons and ions $n_{e,i}$ and velocities of electron and ion fluids $\mathbf{v}_{e,i}$ which take into account thermal fluxes of ions and electrons $\mathbf{q}_i = -\kappa_i \nabla T_i$ and $\mathbf{q}_e = -\kappa_\parallel^e \nabla_\parallel T_e - \kappa_\perp^e \nabla_\perp T_e - \kappa_\wedge^e \mathbf{b} \times \nabla_\perp T_e$. Under condition of frequent collisions tensors of pressure of electrons $P_e$ and ions $P_i$ are reduced to scalar values $P_{e,i} = n_{e,i} T_{e,i}$, and ion and electron temperatures $T_{e,i}$ differ from temperature of neutral component of plasma slightly. It is necessary to consider also large-scale electric field $\mathbf{E} = \mathbf{E}_0 - \nabla \phi$ where $\mathbf{E}_0$ is a background electric current responsible average drift of electrons relatively ion in traversal direction to a magnetic field. The electric potential $\phi$ satisfies to a Poisson equation $\Delta \phi = 4\pi e (n_e - n_i - n_{res})$. Here, according to linear theory of FB instability we separate ions on two groups: the bulk of the ion population with density $n_i$ and group of resonant particles with density $n_{res} \ll |n_i - n_0|$ which are responsible for Landau damping and demand the kinetic description of their motion. In this work we will consider them in linear approximation and in particularly use for them the continuity law $\frac{\partial}{\partial t} n_{res} + \frac{1}{Ze} div \mathbf{j}_{res} = 0$, where $\mathbf{j}_{res}$ is an electric current of resonant ions, $Z$ a charge of ions. The coefficients of thermal conductivities $\kappa_{\parallel,\perp,\wedge}^e$ and are well known and can be readily find in the in the case when electron-ion collisions are more frequent than electron-neutral collision [?], as well in another more realistic case in the lower ionosphere when the collisions of ions and electrons with neutrals are dominated.

In the first step of the model simplification we solve the equation of electron motion in drift approximation which is valid at considered conditions ($\frac{\partial}{\partial t} \ll v_e \ll \omega_e = \frac{eB}{m_e C}$). Introducing a new potential $\Psi = \phi - \frac{\gamma_e T_e}{e} \ln P_e \simeq \phi - \frac{T_e}{e}\left(\frac{\delta T_e}{T_e} + \frac{\delta n_e}{n_e}\right)$, so $\mathbf{E} + \frac{\nabla P_e}{en} = \mathbf{E}_0 - \nabla \Psi$ find the electron velocity $\mathbf{v}_e$ ($\mathbf{z} = \frac{\mathbf{B}}{B}$)

$$v_{ze} = \frac{e}{m_e v_e} \frac{\partial \Psi}{\partial z}, \quad v_{e\perp} \simeq v_D + \frac{c}{B}[\mathbf{z}, \nabla \Psi] - \frac{v_e}{\omega_e}[\mathbf{z}, \mathbf{v}_D] + \frac{c}{B}\frac{v_e}{\omega_e} \nabla_\perp \Psi. \tag{6}$$

Then equation for electron density became

$$\left(\frac{\partial}{\partial t} + \mathbf{v}_d \nabla_\perp\right)\frac{\delta n_e}{n} + \frac{c}{B}[\nabla \ln n_0, \nabla \Psi]_z + \frac{c}{B}\frac{v_e}{\omega_e}\left(\nabla_\perp^2 + \frac{\omega_e^2}{v_e^2}\frac{\partial^2}{\partial z^2}\right)\Psi =$$

$$= \frac{c}{B}\left[\nabla \Psi, \nabla \frac{\delta n_e}{n}\right]_z - \frac{v_e}{\omega_e}\frac{c}{B}\left(\frac{\omega_e^2}{v_e^2}\frac{\partial \Psi}{\partial z}\frac{\partial}{\partial z} + \nabla_\perp \Psi \nabla_\perp\right)\frac{\delta n_e}{n}. \tag{7}$$

Comparing the nonlinear term in (7) one can see that the term $\sim \frac{c}{B}\left[\nabla \frac{\delta n_e}{n}, \nabla \Psi\right]_z$ is the main. The estimations of nonlinear terms in ion motion show that at $\omega \geq \omega_i$ it is possible neglect them and put $\left(\frac{\partial}{\partial t} + \mathbf{v}_i \cdot \nabla\right) \simeq \frac{\partial}{\partial t}$. Indeed, comparing characteristic time of nonlinear ion motion $1\tau_i \sim 1/k|\mathbf{v}_i|$ with characteristic time of electron motion $\tau_e \sim 1/\frac{cm_i}{eB}[\mathbf{k}',\mathbf{k}]_z \frac{e\phi}{m_i}$ one can see that ion times are essentially longer than electrons $\tau_i \geq \tau_e$ if $\frac{\omega}{\omega_i}[\mathbf{k}',\mathbf{k}]_z \geq kk'$. Thus, introducing potential of ion velocity $\mathbf{v}_i = \nabla\Phi$, get fluid equations which describe dynamics of ion bulk component

$$\frac{\partial}{\partial t}\frac{\delta n_i}{n_0} = -\nabla^2\Phi, \nabla^2\left[\left(\frac{\partial}{\partial t}+v_i\right)\Phi + \frac{e\phi}{m_i} + \frac{T_i}{m_i}\left(\frac{\delta T_i}{T_i}+\frac{\delta n_i}{n_0}\right)\right] = 0. \tag{8}$$

Considering oscillations with length of waves essentially exceeding a Debye radii there are every reason to assume their quasineutrality $n_e \simeq n_i = n$, what can be checked easily in the linear approach. If violation of quasi-neutrality neglected $\delta n_e \simeq \delta n_i + n_{res} \simeq \delta n + n_{res}$, a coordination of a motion of electrons of ions $\frac{\partial}{\partial t}(\delta n_e - \delta n_i - n_{res}) = divn(\mathbf{v}_i - \mathbf{v}_e) + \frac{1}{Zen_0}div\mathbf{j}_{res} \simeq 0$ impose restrictions which allow to define potential $\psi$, which is obtained by comparing (8) and (7)

$$\frac{c}{B}\frac{v_e}{\omega_e}\left(\nabla_\perp^2\Psi + \frac{\omega_e^2}{v_e^2}\frac{\partial^2\Psi}{\partial z^2}\right) + \frac{c}{B}[\nabla\Psi, \nabla_\perp \ln n]_z + (\mathbf{v}_D\nabla_\perp)\frac{\delta n}{n} - \frac{v_e}{\omega_e}[\mathbf{v}_D, \nabla_\perp]\frac{\delta n}{n}$$

$$+\frac{c}{B}\left[\nabla\Psi, \nabla_\perp \frac{\delta n}{n}\right] + \frac{c}{B}\frac{v_e}{\omega_e}\left(\nabla_\perp \frac{\delta n}{n}\nabla_\perp \Psi + \frac{\omega_e^2}{v_e^2}\frac{\partial}{\partial z}\frac{\delta n}{n}\frac{\partial \Psi}{\partial z}\right) = \nabla^2\Phi + \frac{1}{Zen_0}div\mathbf{j}_{res}, \tag{9}$$

Note that this condition do not contain derivatives on time and is not evolutionary.
In the short wave region an electron thermal conduction changes properties of oscillations and should be taking into account, but ion thermal conductivity is small and can be neglected. Thus, with variations of electronic entropy $s_e = \delta S_e = \delta \ln \frac{T_e^{3/2}}{n_e}$ ($\frac{\delta T_e}{T_e} = \frac{2}{3}\left(\frac{\delta n_e}{n}+s_e\right)$), from the equation of heat transfer have

$$\left(\frac{\partial}{\partial t}+\mathbf{v}_e \cdot \nabla\right)s_e = v_e\left(\alpha_{ez}\lambda_e^2\nabla_\parallel^2 + \alpha_{e\perp}\rho_e^2\nabla_\perp^2\right)\frac{\delta T_e}{T_e} \tag{10}$$

The full set of equations of FB oscillations (7)-(10) in $k$-space (according $\frac{\delta n}{n} = \sum n_k \exp i\mathbf{kr}$ and with $\psi = \frac{e\Psi}{T_e}$, $\rho_e^2 = \frac{T_e}{m_e\omega_e^2}$, $\varphi = \frac{m_i\Phi}{T_e}$) is

$$\frac{\partial}{\partial t}n_k = c_e^2k^2\varphi_k, \left(\frac{\partial}{\partial t}+v_i\right)\varphi_k + \psi_k + \frac{5}{3}\frac{T_e+T_i}{T_e}n_k = 0 \tag{11}$$

$$i\mathbf{v}_d\mathbf{k}_\perp n_k + i\omega_e\rho_e^2[\nabla \ln n_0, \mathbf{k}]_z\psi_k - v_e\rho_e^2\left(\mathbf{k}_\perp^2 + \frac{\omega_e^2}{v_e^2}k_z^2\right)\psi_k + c_e^2k^2\varphi_k + \left(\frac{div\mathbf{j}_{res}}{Zen_0}\right)_k$$

$$= -\omega_e\rho_e^2\left[\nabla\frac{\delta n}{n}, \nabla\psi\right]_{zk} \tag{12}$$

$$\left(\frac{\partial}{\partial t}+i\mathbf{v}_d\mathbf{k}_\perp\right)s_k=\nu_e\rho_e^2\left(\alpha_{e\perp}\mathbf{k}_\perp^2+\alpha_{ez}\frac{\omega_e^2}{v_e^2}k_z^2\right)\frac{2}{3}(n_k+s_k) \qquad (13)$$

Note that only the third, not evolutionary equation contains nonlinear term. In linear approach excluding amplitudes from above equations (11-12) get the modified dispersion equation

$$\Omega(\Omega+i\nu_i)-c_s^2k^2=$$

$$=c_e^2k^2\left(\frac{-i(\Omega-\mathbf{v}_d\mathbf{k}_\perp)}{\nu_e\rho_e^2\left(\mathbf{k}_\perp^2+\frac{\omega_e^2}{v_e^2}k_z^2\right)-\left(\frac{div\mathbf{j}_{res}}{Zen_0}\right)_k}+\frac{i\frac{2}{3}\frac{2}{3}\nu_e\left(\alpha_{e\perp}\mathbf{k}_\perp^2\rho_e^2+\alpha_{ez}\frac{\omega_e^2}{v_e^2}k_z^2\rho_e^2\right)}{\Omega-\mathbf{v}_d\mathbf{k}_\perp-\frac{2}{3}i\nu_e\left(\alpha_{e\perp}\mathbf{k}_\perp^2\rho_e^2+\alpha_{ez}\frac{\omega_e^2}{v_e^2}k_z^2\rho_e^2\right)}\right) \qquad (14)$$

If to neglect a thermal conduction ($\frac{\delta T_{e,i}}{T_i}=\frac{2}{3}\frac{\delta n}{n}$) and effect of resonant particles one get classical FB model (used for example in simulation [10]). In such case from linear approximation of (11-12) follows dispersion equation which was obtained in basic works [11] and considered. Not difficult to see that the electron thermal conductivity is not essential

$\mathbf{k}\mathbf{v}_d\left(\frac{\Psi_0}{1+\Psi_0}\right)\gg\frac{2}{3}\nu_e\left(\alpha_{e\perp}\mathbf{k}_\perp^2\rho_e^2+\alpha_{ez}\frac{\omega_e^2}{v_e^2}k_z^2\rho_e^2\right)$ and the instability is stabilized at not very short wavelengths $\mathbf{k}_\perp^2\rho_e^2\sim\mathbf{k}\mathbf{v}_d/\nu_e\ll1$.

In modelling of FB instability [6] and [12]on the basis of magnetohydrodynamic quasineutral model was used. Thus they had to apply special efforts to resolve not evolutionary non-linear equation setting communication between fluctuations of density and electric potential. Besides, as it is scored above, considering oscillations in small wavelength it is necessary to take into account Landau damping on ions and a heat conduction by electrons. It is possible to make further simplification of the mathematical description which will maintain all essential lines of interaction of oscillations and can be used for definition of their spectrums. Really, as, according to the linear theory of instability exists a little velocities of increase of oscillations in comparison with a continuance of oscillations it is natural to guess, that at instability stabilization the characteristic return time of non-linear interaction of waves also will be a little in comparison with a continuance of oscillations. Therefore it is obviously possible to search for the solution in weak coupling approach $n_k\simeq n_k(t)\exp(-i\Omega_k t)$, $\frac{\partial}{\partial t}n_k(t)\ll\Omega_k n_k(t)$. The given approach is similar used in the work [13] and allows to reduce essentially computing difficulties reducing a problem to the analysis of combined equations for slowly varying amplitudes of density oscillations.

So, we will substitute $n_k\simeq n_k(t)\exp(-i\Omega_k t)$ in the equations gained from a complete set of equation for density perturbations (11)-(13) in Fourier representation of amplitudes of oscillations. Then with the help of $\psi=\frac{e\Psi}{T_e}=\frac{e\phi}{T_e}-\frac{5}{3}\frac{\delta n_e}{n}-\frac{2}{3}s_e$ and $\phi_k=\psi_k+\frac{5}{3}n_k+\frac{2}{3}s_k$, assuming weak interaction, we can express $\varphi_k$, $\phi_k$ and $s_k$ through $n_k$ approximately, for example as $\varphi_k=\frac{1}{c_e^2k^2}\left(-i\Omega_k+\frac{\partial}{\partial t}\right)n_k$. Then, substituting obtained expressions in equations (11)-(13) and recovering fast phase dependence, we get the final set of equation for amplitudes of density fluctuations

$$\left(\frac{\partial}{\partial t}+i\Omega_k-\gamma_k\right)n_k=\frac{\omega_e}{2}\sum_{\mathbf{k}=\mathbf{k}_1+\mathbf{k}_2}G(k_1,k_2)n_{k_1}n_{k_2} \qquad (15)$$

In derivation of (15), we assume, according the approximation of weak interaction that liner terms annihilates each other in the first approximation. It results that $\Omega_k$ should be find from equation which coincide with expression for real part of dispersion equation (14). At the same time it is possible to show that $\gamma_k$ is given by the solution of equation (14) at $\gamma_k \ll \Omega_k$. Also the small values $\sim \frac{\partial}{\partial t} n_k$ in nonlinear terms should be omitted. For quantities entering into this equation rather cumbersome expressions are obtained, so we present here them in the limit when corrections related to a thermal conduction are negligible

$$G(k_1, k_2) \equiv G(k_2, k_1) \simeq \frac{\rho_e^2 [\mathbf{k}_1, \mathbf{k}_2]_z}{A_k} \left( \frac{\Omega_{k_1}^2}{c_e^2 k_1^2} - \frac{\Omega_{k_2}^2}{c_e^2 k_2^2} \right)$$

The equation features evolution of a spectrum of oscillations of density at development of FB instability taking into account non-linear interaction of waves. Stabilization of FB instability can come, if the effective mechanism of pumping-over of energy from the unstable waves spread in a cone, to strongly decaying waves out of this cone is provided. Such pumping-over can be reached as a result of a cascade of three-wave processes with reduction of a wave length leading to energy pumping-over in field of strong Landau damping on ions. However there is shorter way to transfer energy to damping region, namely if as a result of three-wave processes the unstable wave immediately transmits energy to decaying oscillations. Let us consider a case of three waves which interaction it is featured by combined equations ($\mathbf{k}_1 = \mathbf{k}_2 + \mathbf{k}_3$). High efficiency of interaction of waves is realized if the synchronism requirement is satisfied $\Omega_1 \simeq \Omega_2 + \Omega_3$. However it is not enough of it, as because of the vector nature of nonlinearity its efficiency can be small $G \simeq 0$, if waves are collinear, i.e. in addition it is required, that coefficients $G$ were great. As it is easy to see, necessary requirements are reached for example if $k_{2y} \simeq -k_{3y} \gg k_{1y}$ then $[\mathbf{k}_3, \mathbf{k}_2]_z = (k_{2x} k_{3y} - k_{2y} k_{3x}) \simeq k_{2y}(k_{2x} + k_{3x}) = k_{2y} k_{1x}$ and $[\mathbf{k}_1, \mathbf{k}_2]_z \simeq (k_{2x} k_{1y} - k_{2y} k_{1x}) \simeq k_{2y} k_{1x}$, that is all coefficients of one order and from $k_{2x} + k_{3x} = k_{1x}$ with good accuracy follows mode synchronization.

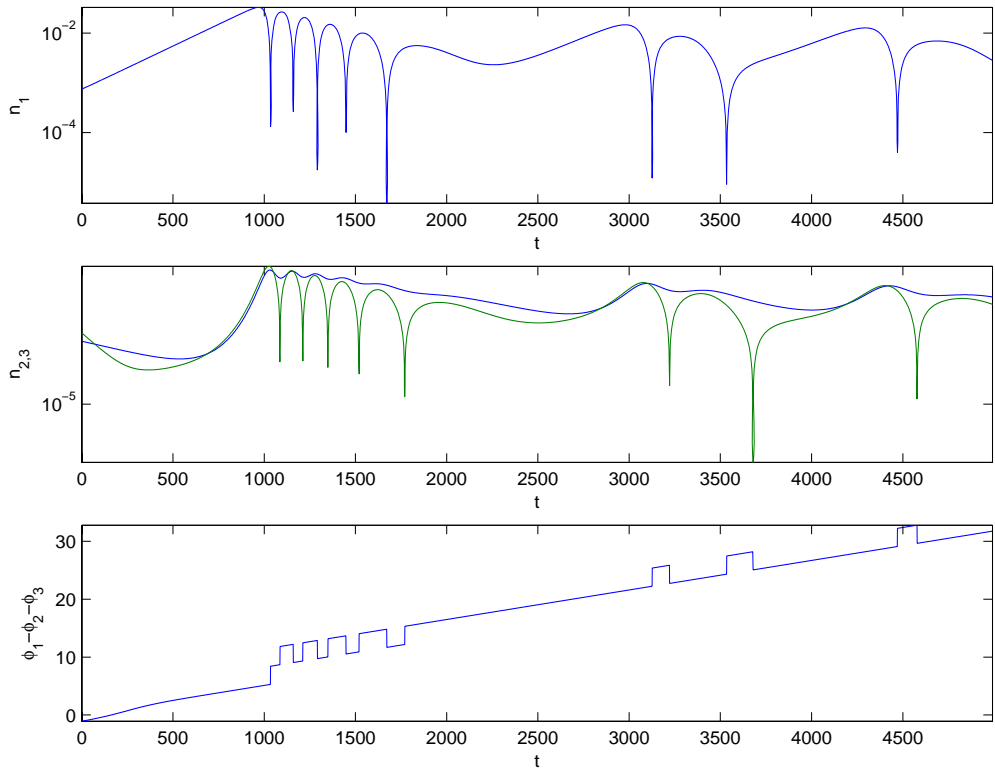

**1. Density amplitude variations for three interacting modes**

One example of three mode dynamics obtained in result of numerical solution of equations in the case when the first mode is unstable and other modes have damping is shown on figure**1**. All variables are presented in normalized values.

**The work was supported by KBN grant OT00A 014 29.**